\documentclass{aa}  

\usepackage{graphicx}
\usepackage{lscape}
\usepackage{longtable}
\usepackage{natbib}
\usepackage{color}
\usepackage{array} 
\bibpunct{(}{)}{;}{a}{}{,} 

\usepackage{txfonts}
\begin{document}

  \title{K2-30\,b and K2-34\,b: two inflated hot-Jupiters \\ around Solar-type stars} 

   \subtitle{}

   \author{
   J.~Lillo-Box\inst{\ref{eso}}\fnmsep\inst{\ref{cab}}, 
   O.~Demangeon\inst{\ref{aix}}, 
   A.~Santerne\inst{\ref{caup}}\fnmsep\inst{\ref{aix}}, 
   S.~C. C. Barros\inst{\ref{caup}}, 
   D.~Barrado\inst{\ref{cab}}, 
   G.~H\'ebrard\inst{\ref{iap}}\fnmsep\inst{\ref{ohp}}, 
   H.~P.~Osborn\inst{\ref{wwick}}, 
   D.~J.~Armstrong\inst{\ref{wwick}}\fnmsep\inst{\ref{belfast}}, 
   J.-M. Almenara\inst{\ref{ipag}}\fnmsep\inst{\ref{cnrs}}, 
   I.~Boisse\inst{\ref{aix}}, 
   F.~Bouchy\inst{\ref{aix}}\fnmsep\inst{\ref{geneve}}, 
   D.~J.~A.~Brown\inst{\ref{wwick}}, 
   B.~Courcol\inst{\ref{wwick}}, 
   M.~Deleuil\inst{\ref{aix}}, 
   E.~Delgado~Mena\inst{\ref{caup}}, 
   R.~F.~D\'iaz\inst{\ref{wwick}}, 
   J.~Kirk\inst{\ref{wwick}}, 
   K.~W.~F.~Lam\inst{\ref{wwick}}, 
   J.~McCormac\inst{\ref{wwick}}, 
   D.~Pollacco\inst{\ref{wwick}}, 
   A.~Rajpurohit\inst{\ref{aix}}\fnmsep\inst{\ref{india}}, 
   J.~Rey\inst{\ref{geneve}}, 
   N.~C.~Santos\inst{\ref{caup}}\fnmsep\inst{\ref{porto}}, 
   S.~G.~Sousa\inst{\ref{caup}}, 
   M.~Tsantaki\inst{\ref{caup}}, 
   P.~A.~Wilson\inst{\ref{iap}}
          }

\institute{European Southern Observatory (ESO), Alonso de Cordova 3107, Vitacura, Casilla 19001, Santiago de Chile (Chile)\label{eso}\\
              \email{jlillobox@eso.org}
\and Depto. de Astrof\'isica, Centro de Astrobiolog\'ia (CSIC-INTA), ESAC campus 28692 Villanueva de la Ca\~nada (Madrid), Spain\label{cab} 
\and Aix Marseille Universit\'e, CNRS, Laboratoire d'Astrophysique de Marseille UMR 7326, 13388, Marseille, France\label{aix} 
\and Instituto de Astrof\' isica e Ci\^encias do Espa\c{c}o, Universidade do Porto, CAUP, Rua das Estrelas, PT4150-762 Porto, Portugal\label{caup}
\and Institut d'Astrophysique de Paris, UMR7095 CNRS, Universit\'e Pierre \& Marie Curie, 98bis boulevard Arago, 75014 Paris, France\label{iap} 
\and Observatoire de Haute-Provence, Universit\'e d'Aix-Marseille \& CNRS, 04870 Saint-Michel l'Observatoire, France\label{ohp}
\and Department of Physics, University of Warwick, Gibbet Hill Road, Coventry, CV4 7AL, UK\label{wwick}
\and Universit\'e Grenoble Alpes, IPAG, 38000 Grenoble, France\label{ipag} 
\and CNRS, IPAG, 38000 Grenoble, France\label{cnrs}
\and Observatoire Astronomique de l'Universit\'e de Gen\`eve, 51 chemin des Maillettes, 1290 Versoix, Switzerland\label{geneve}  
\and Astronomy and Astrophysics Division, Physical Research Laboratory, Ahmedabad 380009, India\label{india}
\and ARC, School of Mathematics \& Physics, Queen's University Belfast, University Road, Belfast BT7 1NN, UK\label{belfast}
\and Depto. de F\'isica e Astronomia, Faculdade de Ci\^encias, Universidade do Porto, Rua Campo Alegre, 4169-007 Porto, Portugal\label{porto}
}
   \date{\today}
   
\titlerunning{K2-30\,b and K2-34\,b: two inflated hot-Jupiters around Solar-type stars}
\authorrunning{Lillo-Box et al.}

 
  \abstract
   {We report the discovery of the two hot-Jupiters K2-30\,b and K2-34\,b. The two planets were detected transiting their main-sequence stars with periods $\sim 4.099$ and $\sim 2.996$~days, in campaigns 4 and 5 of the extension of the \emph{Kepler} mission, K2. Subsequent ground-based radial velocity follow-up with SOPHIE, HARPS-N and CAFE established the planetary nature of the transiting objects. We analysed the transit signal, radial velocity and spectral energy distributions of the two systems to characterize their properties. Both planets (K2-30\,b and K2-34\,b) are bloated hot-Jupiters (1.20 $R_{\rm Jup}$ and 1.22 $R_{\rm Jup}$) around relatively bright ($V =13.5$ and $V=11.5$), slow rotating main-sequence (G8 and F9) stars. Thus, these systems are good candidates for detecting the Rossiter-MacLaughlin effect to measure their obliquity and for atmospheric studies.}

   \keywords{Planets and satellites: detection, gaseous planets; Techniques: radial velocities, high angular resolution, photometric.
               }

   \maketitle
%
\section{Introduction}
The extension of the \emph{Kepler} mission \citep[K2,][]{howell14} is photometrically monitoring different fields along the ecliptic for $\sim$80 day timespans. Despite the shorter timespan and the slightly lower photometric precision with respect to the prime mission, several tens of extrasolar planets have been detected and characterized so far. These planets cover a wide range of properties, from disintegrating Neptune-size objects \citep{sanchis-ojeda15} through validated Earth-size planets \citep[e.g.][]{crossfield15,petigura15} to resonant multi-planetary systems \citep{armstrong15b,barros15}. 

Several works have provided planet candidates based on independent analysis of the light curves \citep[e.g.,][]{foreman-mackey15}. However, as in the prime part of the mission, any candidate requires follow-up observations to unveil its nature. Due to the large \emph{Kepler} pixel size (around 4$\times4$ arcsec), contaminant sources can lie within the photometric aperture. Several high-spatial resolution follow-up surveys were carried out for the prime mission \citep[e.g.][]{lillo-box12,lillo-box14b,adams12,law13} concluding that 20-40\% of the candidates have stellar companions closer than 3 arcsec. Thus,  both high-resolution images and radial velocity (RV) data are needed.

The currently known extrasolar planets show an interesting population of close-in ($a<0.1$~au) gaseous planets, known as hot-Jupiters (hereafter HJ), with a maximum frequency at a 5-day period \citep{santerne16}. The detection and full characterization of these systems has become crucial to understanding early migration processes as well as planet-star and planet-planet interactions.  These are the best targets to study these processes because their consequences are easily detectable from the ground.
For instance, HJs were found to be mostly solitary \citep{steffen12b}, with no other planets in the system. This was explained by possible suppression of rocky planet formation due to the inward migration of the HJ early in the evolution of the system \citep[e.g.,][]{armitage03}. However, the detection of inner and outer planets to the HJ WASP-47\,b using K2 \citep{becker15} has challenged this scenario.
Also, measuring the spin-orbit angle provides hints on the migration history of the system \citep{morton11b} and this angle has been determined for many HJs so far \citep[e.g.,][]{winn05}.  Additionally, many HJs are found to be bloated. The source of such inflation is still not well understood, and several mechanisms have been proposed \citep[e.g.,][]{batygin10,showman02}. But none can explain the inflation by itself. 
On top of this, transiting HJs currently represent the best chance to study exoplanet atmospheres, with high precision light curves also allowing phase curve studies. 

Hence, a full characterization of a large population of HJs is necessary to analyze the migration and formation history of these systems as well as study the planet-planet and planet-star interactions. We have used data from different facilities to identify and characterize the extrasolar planets K2-34\,b and K2-30\,b, two inflated hot-Jupiters around Solar-like stars. In this paper we detail the observations, data reduction, analysis and conclusions about these systems.

\section{Observations and data handling}

\subsection{K2 photometry}
\label{sec:K2}

The star K2-30 (EPIC210957318, 03:29:22.07 +22:17:57.9) was observed by K2 during its campaign 4, between 2015-02-07 and 2015-04-23. K2-34 (EPIC212110888, 08:30:18.91 +22:14:09.3) belongs to field-of-view 5, photometrically monitored by K2 between 2015-04-27 and 2015-07-10. The data was reduced using both the Warwick \citep{armstrong15} and the LAM- K2 \citep{barros15} pipelines. The detrended data (see Tables \ref{tab:k2-318} and \ref{tab:k2-888}) show 1.9\% and 0.8\% {dimmings} every 4.099 and 2.996 days for K2-30 and K2-34, respectively (see Figs.~\ref{fig:rv_epic-318} and ~\ref{fig:rv_epic-888}).

\subsection{High-spatial resolution imaging}

We obtained a high-spatial resolution image of K2-34 {and K2-30} with the instrument AstraLux at the 2.2m telescope in Calar Alto Observatory (Spain). We used the lucky-imaging technique, obtaining 90\,000 frames {(45\,000 frames)}, each with an exposure time of 0.040~s {(0.080~s)} for K2-34 {(K2-30)}. using the maximum gain setting. The images were reduced with the observatory pipeline, which performs basic reduction of the individual frames, selects the frames with the best Strehl ratios \citep{strehl1902}, aligns those frames and combines them to provide a final near-diffraction limited image. In this case, we selected the best 10\% of frames, which translates into an effective exposure time of 360 s. No companion is detected within the sensitivity limits of the images. The sensitivity curve in each case was obtained by following the prescriptions in \cite{lillo-box14}, simulating artificial stars at different positions in the reduced image and different contrast magnitudes and counting how many of them are recovered with a 5$\sigma$ signal-to-noise. In summary, the image would have allowed us to detect companions with contrast magnitudes brighter than 3.5 mag {(3.7 mag)} at 0.5 arcsec, 5.3 mag {(5.0 mag)} at 1 arcsec and 8 mag {(6.0 mag)}  at 2 arcsec {for K2-34 (K2-30}). Since no companion is detected within these limits, we assume that K2-34 {and K2-30} are isolated and that their light curves are not polluted by other sources\footnote{{After the submission of this paper, we became aware of the submitted publication by \cite{hirano16} who found a faint companion to K2-34 at $361.3\pm3.5$ mas with $\Delta m_H=6.19\pm0.11$. This source is below our detection limits but due to its faintness it has a negligible {impact on} the photometric analysis. The maximum contrast for a blended eclipsing binary to be able to mimic the detected transit of K2-34 in the Kepler band would be $\Delta m_{\rm Kep}^{\rm max}=5.2$ mag. Consequently, it is not possible that the close companion is the source of the eclipse. Hence, in practice, we can treat the system as being isolated.}}.

\subsection{High-resolution spectroscopy}

We observed the two transited stars with HARPS-N \citep{cosentino12} at the {Telescopio Nazionale Galileo} (TNG, Spain) and SOPHIE \citep{bouchy13} at the {Observatoire de Haute-Provence} (OHP, France). Two additional epochs for K2-34 were obtained with CAFE \citep{aceituno13} at the 2.2m telescope of the Calar Alto Observatory (CAHA, Spain). The three instruments are fiber-fed high-resolution echelle spectrographs with resolving powers of R$\sim$40\,000 (SOPHIE in the high-efficiency mode), $R=110\,000$ (HARPS-N) and $R=63\,000$ (CAFE) and having no movable pieces. They are located in isolated chambers to improve the stability. SOPHIE and HARPS-N are stabilized in temperature and pressure while these ambient conditions are simply monitored in the case of CAFE to check for possible RV drifts. In the three cases, the data was reduced with the corresponding online pipelines\footnote{For details on the CAFE pipeline see \\ http://www.caha.es/CAHA/Instruments/CAFE/softw.html }. The RV is subsequently computed by determining the weighted cross-correlation function (CCF) between the spectra and a G2V binary mask\footnote{In the case of CAFE, see section 2.3 \cite{lillo-box15b}.}  \citep{baranne96,pepe02}. SOPHIE data were {corrected for} the charge transfer {inefficiency} present in the charge-couple device of the instrument \citep{santerne12}. The RVs were also {corrected for} instrumental drifts using the RV standard star HD\,56124 observed during the same nights and following the prescriptions in \cite{santerne14}. CAFE RVs were also corrected using observations of the same standard star during the nights.

For K2-30, seven epochs were obtained with HARPS-N  during four consecutive nights (4-7 January, 2016) {having a signal-to-noise ratio per pixel at 550 nm (S/N) at the level of 15-30}, and five epochs were obtained with SOPHIE in the subsequent seven nights (8-13 January, 2016) {with S/N of 10-25}. This provides a time span of ten days for this $\sim$ 4-day period planet. For K2-34, we obtained 3 epochs with HARPS-N on 4-7 January 2016 {(S/N of 16-50)},  4 epochs with  SOPHIE on 10-15 January 2016 {(S/N of 27-40)}, and two epochs with CAFE on 24-25 of December 2015 {(S/N of 11-13)}. This encompasses a 22-day time span for this $\sim$3-day period planet. The derived RV values are shown in Tables~\ref{tab:rv-318} and \ref{tab:rv-888}. In both cases, the radial velocity variations show no significant correlation with the bisector values, indicating that they do not originate from blended (undetected) stars. We calculated the projected rotational velocity of the star using the corresponding CCF following \cite{boisse10b}, see Table~\ref{tab:results}.

\section{Results}

\subsection{Stellar properties}
\label{sec:spectral_analysis}

The spectral analysis was performed on the HARPS-N data. For K2-30 the seven spectra were combined, {with the final spectrum reaching a S/N of 56. In the case of } K2-34, we used {a single spectrum} {with S/N of 50}. The spectroscopic parameters were derived with the ARES+MOOG method \citep[see][for details]{sousa14} which is based on the measurement of equivalent widths of iron lines with ARES \citep{sousa15}. This method has been used to derive homogeneous parameters for planet-host stars \citep[e.g.,][]{santos13}. The derived properties ($T_{\rm eff}$, $\log{g}$, and [Fe/H]) for this G8V (K2-30) and F9V (K2-34) {stars} were used as priors for the joint analysis of the data (see Sect.~\ref{sec:joint}) and are provided in Table~\ref{tab:priors}. Additionally, we also determined the lithium abundance for these host stars. We found an abundance of A(Li)$ = 2.16 \pm 0.2$~dex for K2-34 ($T_{\rm eff}=6130\pm50$~K) and an upper limit of A(Li)$ < 0.9$~dex for K2-30 ($T_{\rm eff}=5585\pm38$~K). These abundances provide an estimated lower limit for the age of both targets of 2 Gyr when compared to the abundances of members of the NGC752 \citep{sestito04} or  M67 cluster \citep{pasquini08b}.

\subsection{Joint analysis of the data}
\label{sec:joint}
We used the  PASTIS software \citep{diaz14b,santerne15} to perform a joint analysis of the K2 light curve, radial velocities, and magnitudes of the two targets. The transit signals were modeled\footnote{Models are numerically integrated over the \emph{Kepler} exposure time with an oversampling factor of 10.} using a  modified version of the JKTEBOP code \citep[][and references therein]{southworth11} and {a Keplerian orbit was fitted to the RV data}. The spectral energy distribution (SED, Table~\ref{tab:sed-318}) was modeled with the BT-SETTL library \citep{allard14} and the stellar parameters were derived using the Dartmouth stellar evolution tracks \citep{dotter08}.

\begin{figure}[hbtp]
\includegraphics[width=0.49\textwidth]{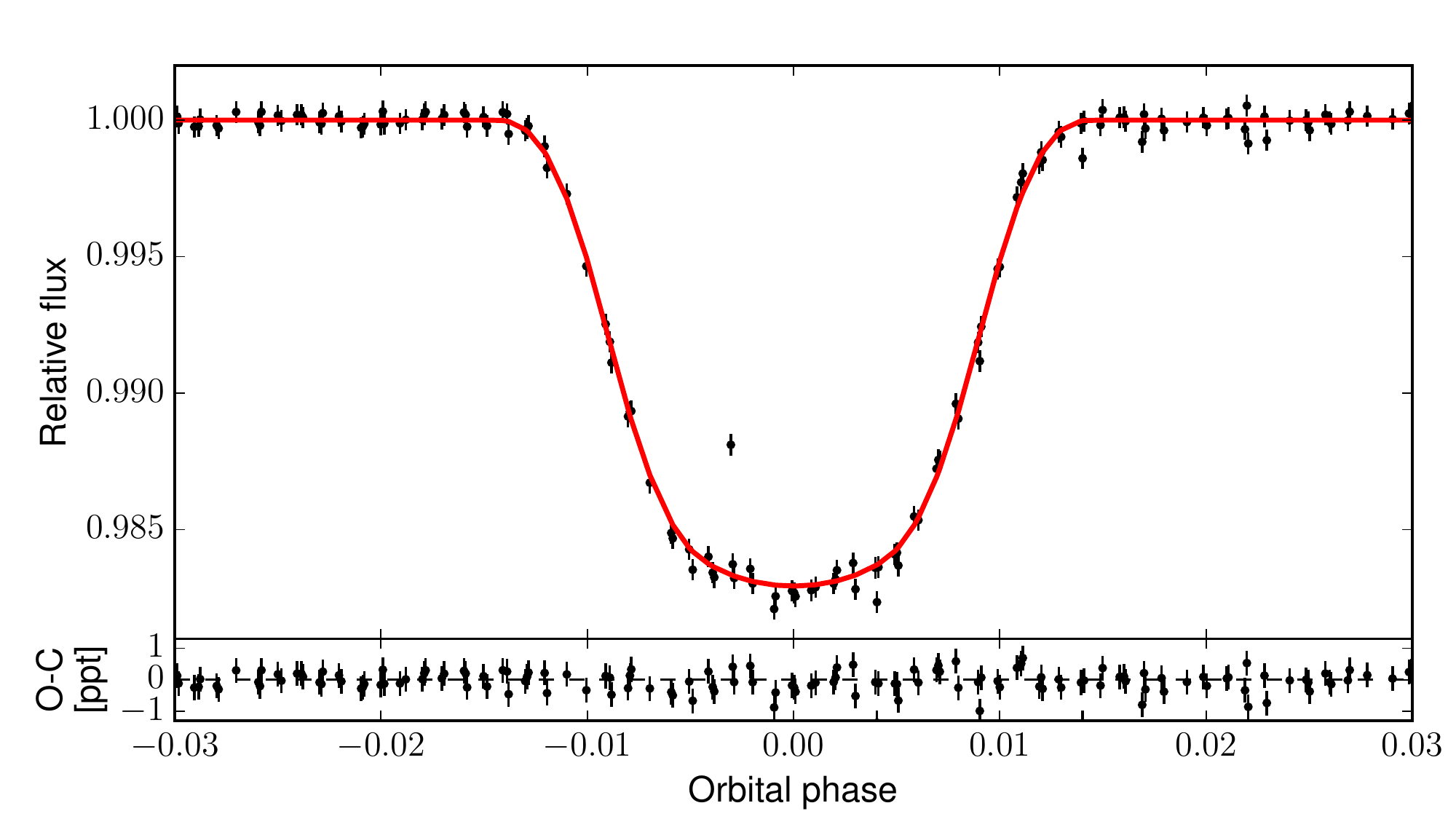}
\includegraphics[width=0.49\textwidth]{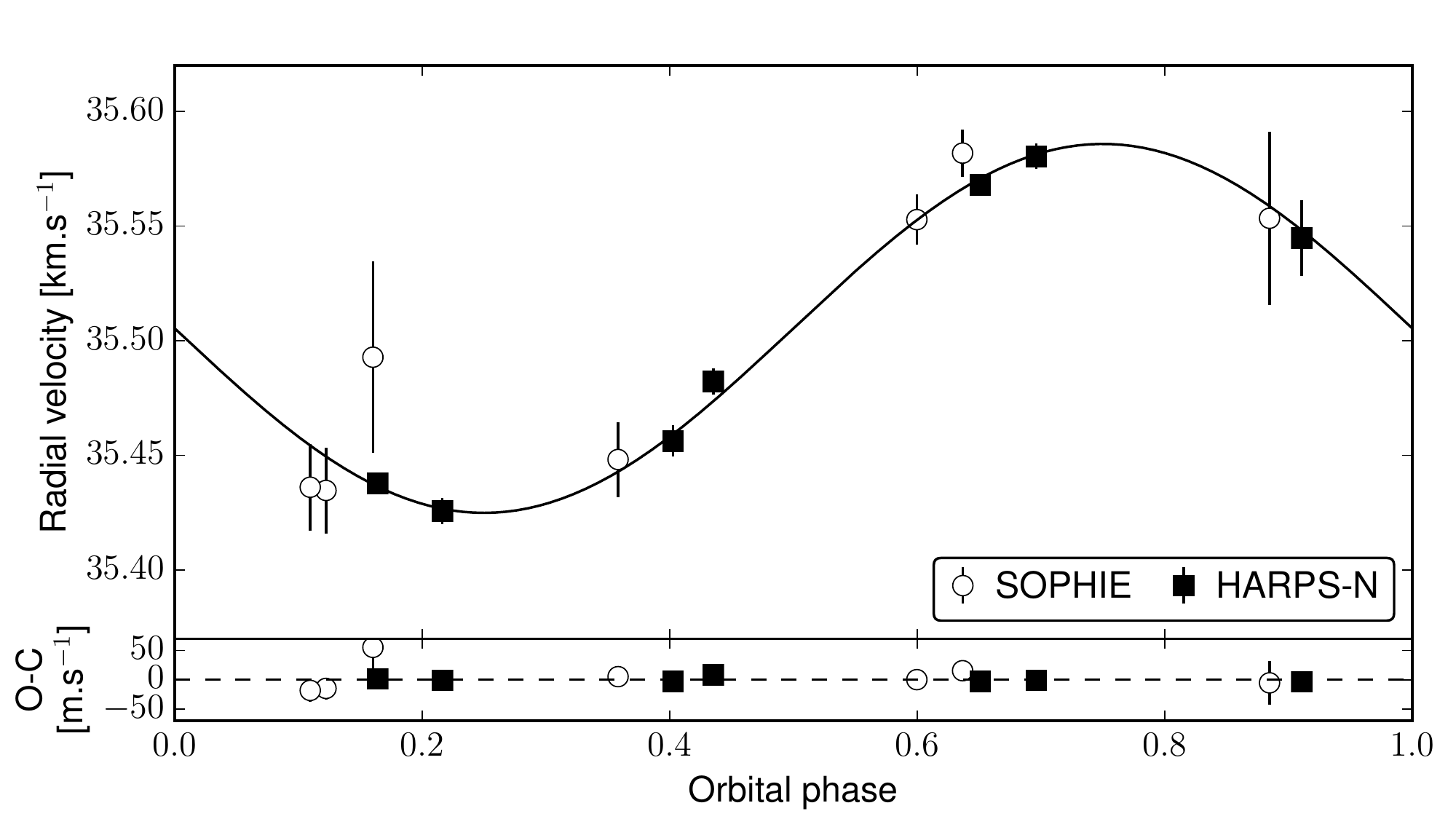}
\caption{Results of the joint analysis with PASTIS for K2-30, including the primary transit (top panel) and radial velocity (bottom panels). The final models are shown with solid lines and the residuals of the data are presented in the lower part of each panel.} 	
\label{fig:rv_epic-318}
\end{figure}

\begin{figure}[hbtp]
\includegraphics[width=0.49\textwidth]{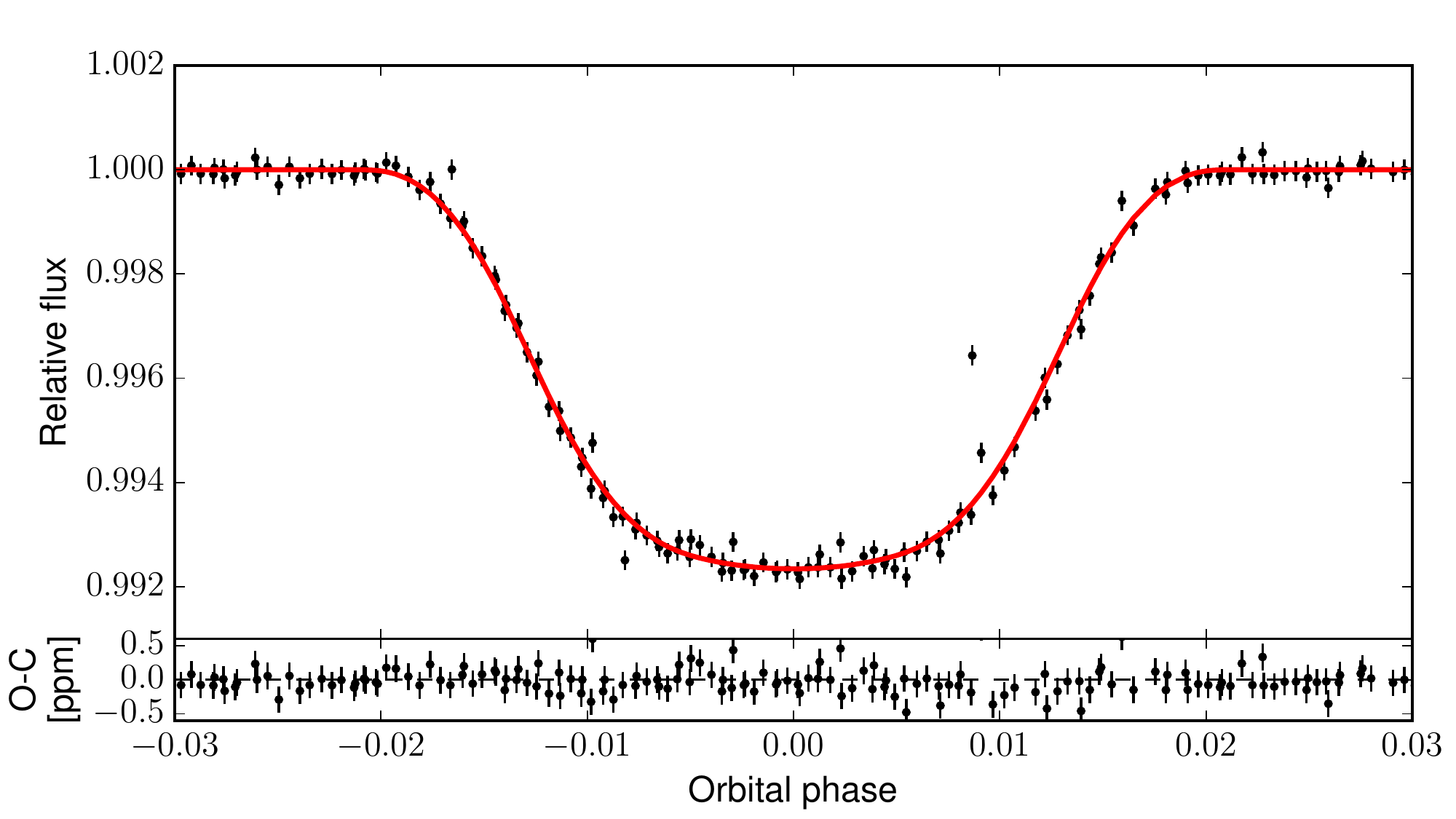}
\includegraphics[width=0.49\textwidth]{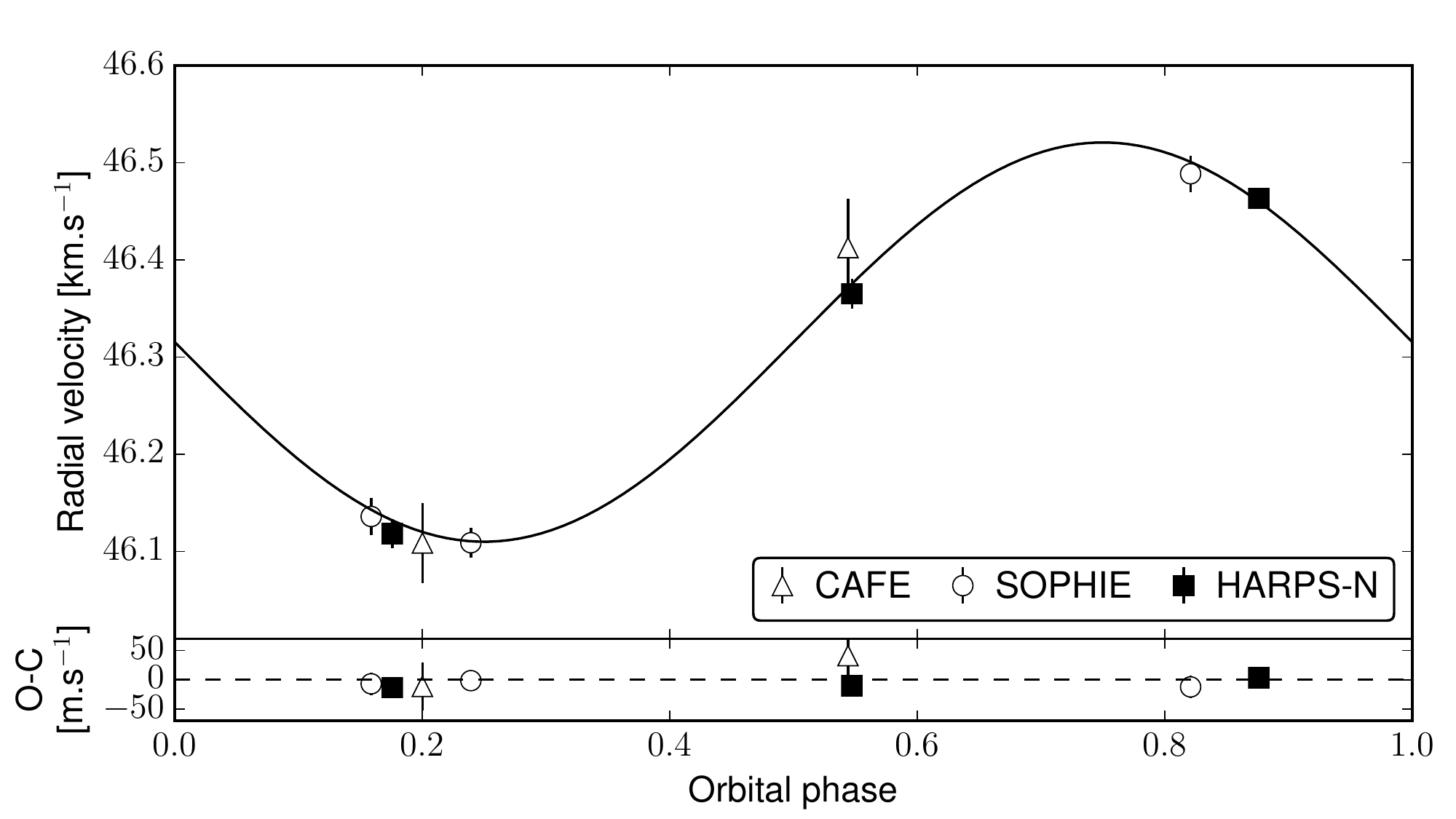}
\caption{Results of the joint analysis with PASTIS for K2-34. Same symbols as in Fig.~\ref{fig:rv_epic-318}.}
\label{fig:rv_epic-888}
\end{figure}

\begin{figure}[hbtp]
\includegraphics[width=0.49\textwidth]{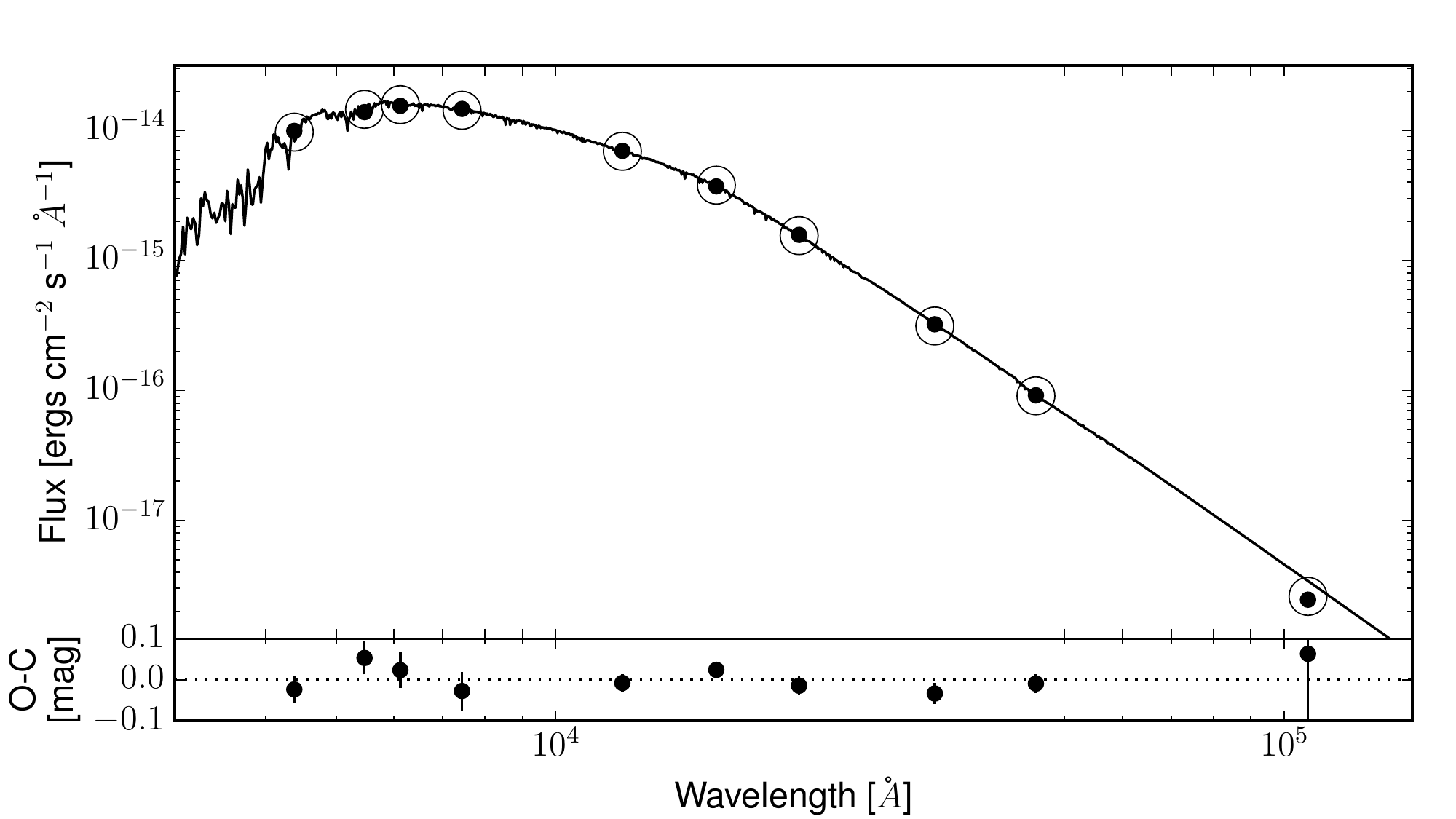}
\includegraphics[width=0.49\textwidth]{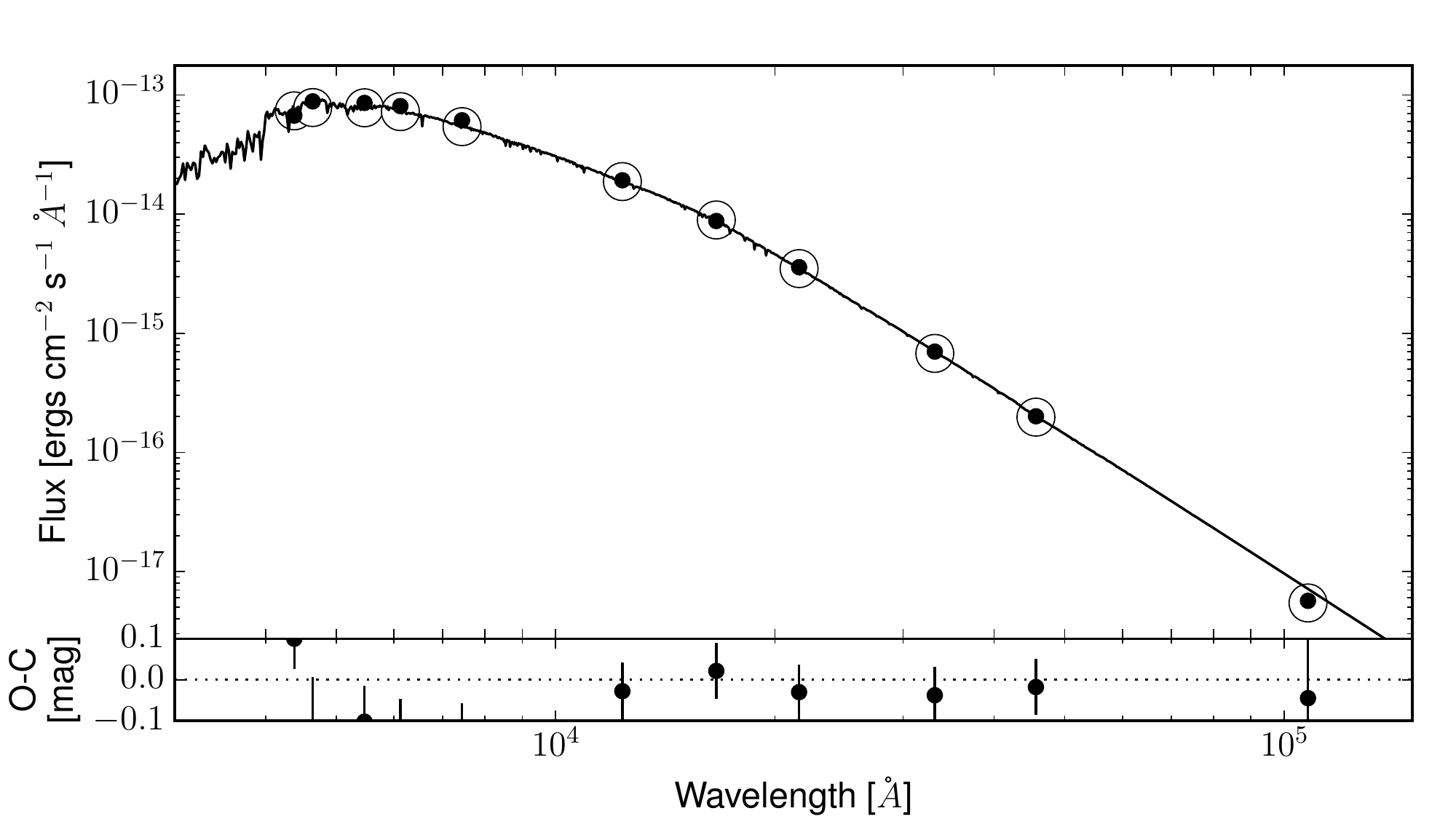}
\caption{Results of the SED fitting in the joint analysis of the data with PASTIS for K2-30 (top panel) and K2-34 (bottom panel). The final models are shown with solid lines and the residuals of the data are presented in the lower part of each panel.} 	
\label{fig:sed}
\end{figure}

We performed a statistical analysis using Markov Chain Monte-Carlo (MCMC) algorithms. The model is described by six free parameters for the host star ($T_{\rm eff}$, $\log{g}$, [Fe/H], systemic radial velocity $V_{\rm sys}$, interstellar extinction $E(B-V)$, and distance $d$) and seven free parameters for the transit (period $P$, epoch of first transit $T_0$, radial velocity amplitude $K$, radius ratio $R_p/R_{\star}$, orbital eccentricity $e$, inclination $i$, and argument of periastron $\omega$). For the light curve, we also fit for an additional source of white noise, the out-of-transit flux level, and the level of contamination. For the RV, we additionally fit for a jitter term for each instrument and a RV offset between SOPHIE and the other instruments used. A jitter is also added for the SED analysis. For K2-34, due to the low number of RV datapoints, we assumed a circular orbit\footnote{{We tested the non-circular hypothesis and compared the bayesian information criterion \citep[BIC, see, for instance,][]{schwarz78,smith09} of the two models. As expected, the results provide strong evidence for the circular model given the current data, with a BIC difference of 2 in favor of the circular (simpler) hypothesis.}}. This assumption is justified by the short period and circularization mechanisms. In total, 20 free parameters are fitted for both systems. Uniform priors are used for all free parameters except for the stellar values that were constrained to the results of the spectral analysis (see Sect.~\ref{sec:spectral_analysis}). The list of priors is shown in Table~\ref{tab:priors}. We ran 20 chains of 3$\times10^5$ iterations randomly drawn from the joint prior distribution. All chains converged toward the same solution which is assumed to be the global maximum. In order to obtain the final solution and its uncertainties, we removed the burn-in phase of each chain and thinned them by computing their maximum correlation length. At this point we merged all of them to compute a well-sampled and clean posterior distribution having more than 2000 independent samples. The median values are presented in Table~\ref{tab:results} together with the 15.7\% and 84.3\% percentiles of the marginalized distributions. In the table we also present other parameters derived from those fitted by the model (e.g., planet mass, planet radius, stellar mass, etc.). The data and best fit models are presented in Figs~\ref{fig:rv_epic-318}, \ref{fig:rv_epic-888} and \ref{fig:sed}.

\section{Discussion}

The combination of different datasets for K2-30\,b and K2-34\,b establishes the planetary nature of the transiting objects around two main-sequence stars (G8V and F9V, respectively) observed by K2. The analysis of the data indicates that both systems are composed of single giant (1.197$\pm$0.052 $R_{\rm Jup}$ for K2-30 and 1.217$\pm$0.053 $R_{\rm Jup}$ for K2-34) planets in close-in orbits around their main-sequence stars.

In the case of K2-30, the insolation flux received by the planet is $F=4.24\pm 0.37 \times 10^{8}$~erg~s$^{-1}$~cm$^{-2}$. This is significantly larger than the empirical cut off limit derived by \cite{demory11} for a planet to have an inflated radius ($F>2.08\times 10^8$~erg~s$^{-1}$~cm$^{-2}$). Assuming a Bond albedo $A=0$ and a complete redistribution of the tidal heating along the planet, its expected radius according to Eq.~9 in \cite{enoch12} would be $0.777\pm0.014$~R$_{\rm Jup}$\footnote{Larger albedos would imply smaller expected planetary radius so this can be considered as an upper limit.}. According to \cite{weiss13} the mass-radius-insolation flux relation provides an expected radius for this planet of $1.151\pm0.010$~R$_{\rm Jup}$.  In the case of K2-34\,b, the planet receives an insolation flux of  $F=1.768 \pm 0.14\times 10^{9}$~erg~s$^{-1}$~cm$^{-2}$. This is one order of magnitude larger than the above mentioned cut off limit derived by \cite{demory11}. The expected radius according to \cite{enoch12} would be $0.993\pm{0.016}$~R$_{\rm Jup}$ and the expected radius from \cite{weiss13} is $1.267\pm{0.010}$~R$_{\rm Jup}$.

\begin{figure}[hbtp]
\includegraphics[width=0.45\textwidth]{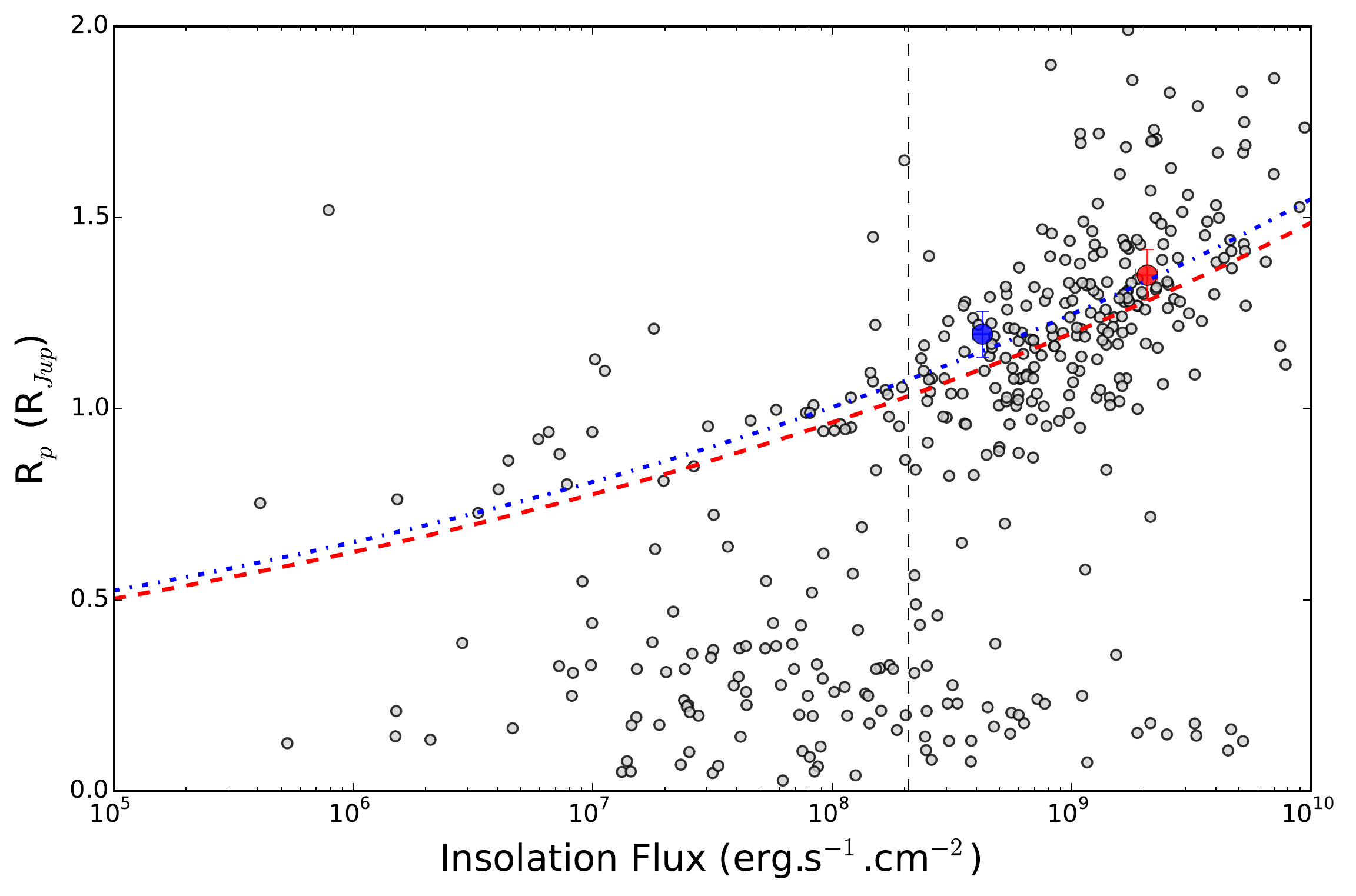}
	\caption{Insolation flux and planet radius for  all extrasolar planets with known radius, $T_{\rm eff}$ and semi-major axis. Data taken from the The Extrasolar Planet Encyclopedia. The two hot-Jupiters found in this work are marked as filled circles (K2-34 in red and K2-30 in blue). The expected radius-insolation flux dependency for each system according to \cite{weiss13} is shown as dashed red (K2-34) and dash-dot blue (K2-30) lines.} 	
	\label{fig:insolation}
\end{figure}

In both cases, the planets have comparable radii to that predicted by empirical calibrations (see Fig.~\ref{fig:insolation}). Although compared to the values predicted by \cite{weiss13} they are compatible within $1\sigma$ with being inflated due to the high stellar insolation flux, both are clearly larger ($>3\sigma$) than the predicted value by \cite{enoch12}. It is known that high stellar irradiance can explain the inflation of the close-in Jupiter planets with radii up to $\sim1.2$~$R_{\rm Jup}$ \citep{guillot02}. However, this cannot explain larger radii, and so other mechanisms must play a role \citep[e.g.,][]{bodenheimer01,batygin10,chabrier07}. The small {(statistically not significant)} eccentricity found in K2-30 could possibly indicate some tidal heating, but other mechanisms cannot be rejected. 

In this paper, we have characterized two HJs. They show bloated radii possibly due to the large stellar insolation that they are receiving from their host. However, other possible mechanisms such as tidal heating may be playing a role. Since the hosts are bright ($V=11.5$ for K2-34 and $V=13.5$ for K2-30) and the planets are inflated, they are amenable for atmospheric characterization either from the ground or from space. As they transit, they are also good candidates for the detection of the Rossiter-MacLaughlin effect to infer the spin-orbit angle and study the evolutionary history of these systems. From the derived parameters, the amplitude of this effect should be around 60 m/s and 40 m/s for K2-30 and K2-34, respectively. Future searches for additional bodies will be interesting to unveil planet-planet interactions during the onset of the planetary systems. \\

\noindent {Note: After the submission of this paper, we become aware of two other publications related to the two targets analyzed in this work. \cite{hirano16} analyzed the K2-34  system, finding good agreement in the parameters directly derived from modeling the observations. Due to small differences in the determined stellar properties, some absolute physical and orbital parameters disagree by a small percentage. \cite{johnson16} analyzed K2-30 and their results are mainly in agreement with ours. Some parameters are still different by a small percentage but this could be due to their assumption of a circular orbit as well as a different approach to the calculation of the stellar parameters.}

\begin{acknowledgements}
J.L-B acknowledges financial support from the Marie Curie Actions of the European Commission (FP7-COFUND) and the Spanish grant AYA2012- 38897-C02-01. O.D. acknowledges support by CNES through contract 567133. A.S. is supported by the European Union under a Marie Curie Intra-European Fellowship for Career Development with reference FP7-PEOPLE-2013-IEF, number 627202. D.J.A. and D.P acknowledge funding from the European Union Seventh Framework programme (FP7/2007- 2013) under grant agreement No. 313014 (ETAEARTH). J-M.A. acknowledges funding from the European Research Council under the ERC Grant Agreement n. 337591-ExTrA. P.A.W acknowledges the support of the French Agence Nationale de la Recherche (ANR), under program ANR-12-BS05-0012 "Exo-Atmos".  AS, SCCB, EDM, NCS, SS, and MT acknowledge support by Funda\c{c}\~ao para a Ci\^encia e a Tecnologia (FCT) through the research grants UID/FIS/04434/2013 (POCI-01-0145-FEDER-007672) and project PTDC/FIS-AST/1526/2014. NCS, SGS, and SCCB acknowledge the support from FCT through Investigador FCT contracts of reference IF/00169/2012, IF/00028/2014, and IF/01312/2014, respectively, and POPH/FSE (EC) by FEDER funding through the program "Programa Operacional de Factores de Competitividade - COMPETE". EDM acknowledges the support from FCT in the form of the grant SFRH/BPD/76606/2011. 
This publication is based on observations collected with the NASA satellite Kepler, with the SOPHIE spectrograph at OHP (CNRS, France), HARPS-N (La Palma, Spain), and CAFE and AstraLux at Calar Alto Observatory (Spain). This publication makes use of data products from the Wide-field Infrared Survey Explorer, which is a joint project (UCLA/JPL) funded by NASA. This research made use of the AAVSO Photometric All-Sky Survey, funded by the Robert Martin Ayers Sciences Fund. 
\end{acknowledgements}


\bibliographystyle{aa} 
\bibliography{../../../biblio2} 

\begin{table}
\setlength{\extrarowheight}{2pt}
\small
\caption{Normalized detrended K2 light curve of K2-30 (EPIC\,210957318, see Sect.~\ref{sec:K2}). The complete version of this table is available online in the CDS.}             
\label{tab:k2-318}      
\centering          
\begin{tabular}{r c c}     
\hline\hline       

BJD-2400000 (days) & Flux & Flux unc \\ \hline
2457067.684383  &  1.000706   &  0.000103  \\
2457067.704815  &  0.999879   &  0.000103  \\
2457067.725247  &  0.999752   &  0.000103  \\
2457067.745679  &  0.999817   &  0.000103  \\
2457067.766111  &  0.999762   &  0.000103  \\
...           &  ...        &  ...       \\

\hline                  
\end{tabular}

\end{table} 

\begin{table}
\setlength{\extrarowheight}{2pt}
\small
\caption{Normalized detrended K2 light curve of K2-34  (EPIC\,212110888, see Sect.~\ref{sec:K2}). The complete version of this table is available online in the CDS.}             
\label{tab:k2-888}      
\centering          
\begin{tabular}{r c c}     
\hline\hline       

BJD (days) & Flux & Flux unc \\ \hline
2457139.630908  & 0.999991  & 0.000042 \\
2457139.651340  & 0.999988  & 0.000042 \\
2457139.671772  & 0.999930  & 0.000042 \\
2457139.692204  & 0.999910  & 0.000042 \\
2457139.712636  & 1.000159  & 0.000042 \\
...           &  ...        &  ...       \\

\hline                  
\end{tabular}

\end{table} 

\begin{table}
\setlength{\extrarowheight}{2pt}
\scriptsize
\caption{Radial velocity data for K2-30.}             
\label{tab:rv-318}      
\centering          
\begin{tabular}{r c c c c}     
\hline\hline       

MBJD (days)  &  RV (km/s) & BIS (m/s) & FWHM (km/s) & Instrument \\ \hline

392.35938  &  35.5624 $\pm$ 0.0043  &  -39.8 $\pm$  6.5  &  6.8842 $\pm$ 0.0087  & HARPS-N\\
392.57382  &  35.5504 $\pm$ 0.0055  &  -18.9 $\pm$  8.2  &  6.8885 $\pm$ 0.0109  & HARPS-N\\
393.33790  &  35.5809 $\pm$ 0.0066  &   -3.2 $\pm$ 10.0  &  6.8780 $\pm$ 0.0133  & HARPS-N\\
393.47098  &  35.6069 $\pm$ 0.0056  &  -37.6 $\pm$  8.4  &  6.8795 $\pm$ 0.0111  & HARPS-N\\
394.35486  &  35.6925 $\pm$ 0.0042  &  -32.3 $\pm$  6.4  &  6.8701 $\pm$ 0.0085  & HARPS-N\\
394.54090  &  35.7050 $\pm$ 0.0054  &  -45.8 $\pm$  8.1  &  6.8717 $\pm$ 0.0108  & HARPS-N\\
395.41956  &  35.6694 $\pm$ 0.0165  &    3.9 $\pm$ 24.7  &  6.8693 $\pm$ 0.0329  & HARPS-N\\

396.44205  &  35.493 $\pm$ 0.042  &  -24.5 $\pm$ 75.1  &  9.682 $\pm$ 0.104  & SOPHIE\\
398.39484  &  35.582 $\pm$ 0.010  &  -35.3 $\pm$ 18.2  &  9.523 $\pm$ 0.025  & SOPHIE\\
399.41136  &  35.554 $\pm$ 0.038  &  -18.1 $\pm$ 67.9  &  9.415 $\pm$ 0.094  & SOPHIE\\
400.38519  &  35.435 $\pm$ 0.019  &  -47.5 $\pm$ 33.5  &  9.517 $\pm$ 0.046  & SOPHIE\\
401.35244  &  35.448 $\pm$ 0.016  &  -22.0 $\pm$ 29.3  &  9.572 $\pm$ 0.040  & SOPHIE\\
402.34190  &  35.553 $\pm$ 0.011  &   -2.7 $\pm$ 19.6  &  9.540 $\pm$ 0.027  & SOPHIE\\
404.43149  &  35.436 $\pm$ 0.019  &  -11.0 $\pm$ 34.0  &  9.591 $\pm$ 0.047  & SOPHIE\\

\hline                  
\end{tabular}
\tablefoot{MBJD = Modified Barycentric Julian Date (BJD-2457000)}
\end{table} 




\begin{table}
\setlength{\extrarowheight}{2pt}
\scriptsize
\caption{Radial velocity data for K2-34.}             
\label{tab:rv-888}      
\centering          
\begin{tabular}{r c c c c}     
\hline\hline       

MBJD (days)  &  RV (km/s) & BIS (m/s) & FWHM (km/s) & Instrument \\ \hline
381.59795  & 45.863 $\pm$ 0.040 & -70 $\pm$  65  & 10.597 $\pm$ 0.080 &      CAFE \\
382.62804  & 46.167 $\pm$ 0.049 & -178 $\pm$  70 & 10.841 $\pm$ 0.110 &      CAFE \\

392.60905  &  46.5458 $\pm$ 0.0034  &  35.3 $\pm$ 5.1   &  9.5015 $\pm$ 0.0069  & HARPS-N\\
393.50729  &  46.2009 $\pm$ 0.0133  &  40.7 $\pm$ 19.9  &  9.4696 $\pm$ 0.0266  & HARPS-N\\
394.61995  &  46.4477 $\pm$ 0.0138  &  36.6 $\pm$ 20.8  &  9.4864 $\pm$ 0.0277  & HARPS-N\\

398.43529  &  46.490 $\pm$ 0.018  &  11 $\pm$ 32  &  11.395 $\pm$ 0.044  & SOPHIE\\
399.44751  &  46.136 $\pm$ 0.018  & -17 $\pm$ 32  &  11.437 $\pm$ 0.045  & SOPHIE\\
399.68876  &  46.109 $\pm$ 0.014  &  51 $\pm$ 25  &  11.554 $\pm$ 0.035  & SOPHIE\\
403.49977  &  46.328 $\pm$ 0.019  &  16 $\pm$ 33  &  11.367 $\pm$ 0.046  & SOPHIE\\

\hline                  
\end{tabular}
\tablefoot{MBJD = Modified Barycentric Julian Date (BJD-2457000)}
\end{table}

\begin{table*}[]
\caption{List of free parameters used in the \texttt{PASTIS} analysis of the light curves, radial velocities and SED with their associated prior. }
\begin{center}
\begin{tabular}{lcc}
\hline
\hline
Parameter & K2-30 & K2-34\\
\hline
\multicolumn{3}{l}{\it Orbital parameters}\\
&&\\
Orbital period $P$ [d] & $\mathcal{N}(4.09803;1\times10^{-3})$ & $\mathcal{N}(2.996;1\times10^{-3})$\\
Epoch of first transit T$_{0}$ [BJD$_{\rm TDB}$] - 2450000 & $\mathcal{N}(7063.826;0.1)$ & $\mathcal{N}(7141.35;0.1)$\\
Orbital eccentricity $e$ & $\beta(0.867;3.03)$ & 0 (fixed) \\
Argument of periastron $\omega$ [\degr] & $\mathcal{U}(0;360)$ & -- \\
Inclination $i$ [\degr] & $\mathcal{S}(70;90)$ & $\mathcal{S}(70;90)$\\
&&\\
\hline
\multicolumn{3}{l}{\it Planetary parameters}\\
&&\\
Radial velocity amplitude $K$ [m/s] & $\mathcal{U}(0;1000)$ & $\mathcal{U}(0;1000)$\\
Planet-to-star radius ratio $a/R_{\star}$ & $\mathcal{U}(0; 0.5)$ & $\mathcal{U}(0; 0.5)$\\
&&\\
\hline
\multicolumn{3}{l}{\it Stellar parameters}\\
&&\\
Effective temperature $T_{\rm eff}$ [K] & $\mathcal{N}(5573;38)$ & $\mathcal{N}(6139;50)$ \\
Surface gravity $\log{g}$ [g/cm$^{2}$] & $\mathcal{N}(4.32;0.20)$ & $\mathcal{N}(4.07;0.21)$\\
Iron abundance [Fe/H] [dex] & $\mathcal{N}(0.14;0.03)$ & $\mathcal{N}(0.11;0.04)$\\
Reddening E(B-V) [mag] & $\mathcal{U}(0;1)$ & $\mathcal{U}(0;1)$\\
Systemic radial velocity V$_{\rm sys}$ [km/s] & $\mathcal{U}(-100, 100)$ & $\mathcal{U}(-100, 100)$\\
Distance to Earth $d$ [pc] & $\mathcal{P}(2;10;1000)$ & $\mathcal{P}(2;10;1000)$\\
Linear limb darkening coefficient ua & $\mathcal{U}(0, 1.2)$ & $\mathcal{U}(0, 1.2)$\\
Quadratic limb darkening coefficient ub & $\mathcal{U}(0, 1.2)$ & $\mathcal{U}(0, 1.2)$\\
&&\\
\hline
\multicolumn{3}{l}{\it Instrumental parameters}\\
&&\\
CAFE radial velocity jitter $\sigma_{\rm RV,CAFE}$ [m/s] & -- & $\mathcal{U}(0;100)$ \\
SOPHIE radial velocity jitter $\sigma_{\rm RV,SOPHIE}$ [m/s] & $\mathcal{U}(0;100)$ & $\mathcal{U}(0;100)$\\
HARPS-N radial velocity jitter $\sigma_{\rm RV,HARPS-N}$ [m/s] & $\mathcal{U}(0;100)$ & $\mathcal{U}(0;100)$\\
CAFE -- SOPHIE radial velocity offset $\Delta$ RV$_{S,CAFE}$ [m/s] & $\mathcal{U}(-1000;1000)$ & $\mathcal{U}(-1000;1000)$\\
HARPS-N -- SOPHIE radial velocity offset $\Delta$ RV$_{S,H-N}$  [m/s] & $\mathcal{U}(-1000;1000)$ & $\mathcal{U}(-1000;1000)$\\
SED jitter $\sigma_{\rm SED}$ [mag] & $\mathcal{U}(0;1)$ & $\mathcal{U}(0;1)$\\
\hline
\hline
\end{tabular}
\tablefoot{$\mathcal{N}(\mu;\sigma^{2})$ is a normal distribution with mean $\mu$ and width $\sigma^{2}$, $\mathcal{U}(a;b)$ is a uniform distribution between $a$ and $b$, $\mathcal{S}(a,b)$ is a sine distribution between $a$ and $b$, $\beta(a;b)$ is a Beta distribution with parameters $a$ and $b$, and $\mathcal{P}(n;a;b)$ is a power-law distribution of exponent $n$ between $a$ and $b$.}
\tablebib{The choice of prior for the orbital eccentricity is described in \citet{kipping13}.}
\end{center}
\label{tab:priors}
\end{table*}%

\begin{table}
\setlength{\extrarowheight}{2pt}
\small
\caption{Photometry used to for the spectral energy distribution analysis of  K2-30 and K2-34}             
\label{tab:sed-318}      
\centering          
\begin{tabular}{r c c}     
\hline\hline       

Band	   &	K2-30	    & K2-34                \\ \hline
Johnson-V  &	13.53	$\pm$ 0.039 & 11.548$\pm$0.057          \\
Johnson-B  &	14.506	$\pm$ 0.030 & 12.429$\pm$0.033            \\
g$^{\prime}$  &    -                   & 11.892$\pm$0.119     \\
r$^{\prime}$  &	13.184	$\pm$ 0.042 & 11.389$\pm$0.026       \\
i$^{\prime}$  &	12.819	$\pm$ 0.046 & 11.264$\pm$0.038        \\
2MASS-J	   &	11.632	$\pm$ 0.019 & 10.528$\pm$0.025    	 \\
2MASS-H	   &	11.190	$\pm$ 0.016 & 10.258$\pm$0.022    	 \\
2MASS-Ks   &	11.088	$\pm$ 0.020 & 10.193$\pm$0.017    	 \\
WISE-W1	   &	11.016	$\pm$ 0.023 & 10.174$\pm$0.023    	 \\
WISE-W2	   &	11.058	$\pm$ 0.021 & 10.207$\pm$0.020    	 \\
WISE-W3	   &	11.067	$\pm$ 0.161 & 10.159$\pm$0.169      \\

\hline                  
\end{tabular}
\tablefoot{Optical magnitudes (Johnson-V, Johnson-B, g$^{\prime}$, r$^{\prime}$, and i$^{\prime}$) were obtained from the APASS database. Infrared values were obtained from \cite{cutri13}.}
\end{table} 

\begin{table}
\setlength{\extrarowheight}{3pt}
\small
\caption{Host star, planet and orbital parameters inferred from the joint analysis of the data. Uncertainties represent the 15.7\% and 84.3\% percentiles of the marginalized posterior distribution}
\label{tab:results}      
\centering          
\begin{tabular}{r c c}     
\hline\hline       

Parameter  &  \textbf{K2-30}  &  \textbf{K2-34} \\ \hline

\multicolumn{3}{l}{\textbf{Planet properties}} \\ \hline
K (m/s)                              &   79.9 $\pm$ 4.0                 & 209 $\pm$ 12         \\
$R_p/R_{\star}$                      &  0.1302 $\pm$ 0.0026             & 0.0905 $\pm$ 0.0017          \\
         $M_p$ ($M_{\rm Jup}$)       &     $0.623 \pm 0.031$            &        $1.76 \pm 0.13$       \\
         $R_p$ ($R_{\rm Jup}$)       &        $1.196 \pm 0.060$         &        $1.350 \pm 0.067$    \\
   $\rho_p$ ($\rho_{\rm Jup}$)       &     $0.364 \pm 0.058$            &        $0.71\pm 0.11$    \\ 
   T$_{\rm eq}$ (K)                  &      $1185^{+44}_{-28}$          &         $1742\pm 38$    \\

\multicolumn{3}{l}{\textbf{Orbital properties}} \\ \hline
Period (days)                        &  4.098507 $\pm$ 0.000028         & 2.995607 $\pm$ 0.000017         \\
T$_0$ (MBJD\tablefootmark{a}, days)  &   63.80710 $\pm$ 0.00027         & 141.35132 $\pm$ 0.00022       \\
i (deg)                              &  86.32 $\pm$ 0.38                & $82.06\pm 0.51$          \\
e                                    &  $0.027^{+0.036}_{-0.020}$       & 0.0 (assumed)          \\
$\omega$ (deg)                       &  $120^{+100}_{-51}$              & -         \\ 
                      $a$ (au)       &  $0.04986 \pm 0.00035$           &   $0.04419^{+0.00063}_{-0.00110}$     \\
             $a/R_{\star}$           &  $11.40\pm0.52$                  &   $6.20\pm0.25$     \\
             b$_{\rm primary}$       &        $0.721 \pm 0.036$         &        $0.857\pm 0.023$    \\
           T$_{\rm dur}$ (hours)     &  $2.355\pm 0.030$                &   $2.492\pm 0.023$ \\  \hline

\multicolumn{3}{l}{\textbf{Host properties}} \\ \hline
     $M_{\star}$ ($M_{\odot}$)       &     $0.984 \pm 0.020$            &        $1.281^{+0.055}_{-0.091}$    \\ 
     $R_{\star}$ ($R_{\odot}$)       &     $0.941\pm 0.041$             &        $1.526\pm 0.076$    \\         
     $\rho_{\star}$ ($\rho_{\odot}$) & $1.18\pm 0.15$                   &        $0.356\pm 0.044$  \\
$V_{\rm sys}$ (km/s)                 &  35.5090 $\pm$ 0.0087              & 46.311 $\pm$ 0.013          \\
d (pc)                               & 323 $\pm$ 14                     & 377 $\pm$ 20        \\
E(B-V)                               & 0.253 $\pm$ 0.014                & 0.036 $\pm$ 0.020         \\
$\log{g}$ (cgs)                      & 4.484 $\pm$ 0.042                & 4.173 $\pm$ 0.032         \\
T$_{\rm eff}$                        & 5581 $\pm$ 38                    & 6132 $\pm$ 47          \\
$[$Fe/H$]$ (dex)                     & 0.136 $\pm$ 0.029                & 0.24 $\pm$ 0.22            \\
$\log{L/L_{\odot}}$ (dex)            & -0.112 $\pm$ 0.039               &    $0.471\pm0.048$        \\
ua                                   & 0.24 $\pm$ 0.18                  & $0.18^{+0.20}_{-0.13}$            \\
ub                                   & 0.34 $\pm$ 0.28                 & 0.28 $\pm$ 0.28              \\ 
$v\sin{i}$ (km/s)                    & $3.9 \pm 1.1$                    & $5.2 \pm 1.4$                \\ \hline

\multicolumn{3}{l}{\textbf{Jitters ($\sigma$) and instrumental offsets $(\Delta RV)$}}\\ \hline 
$\sigma_{\rm RV,HARPS-N}$ (m/s)      & $3.6^{+4.8}_{-2.5}$              & $16^{+29}_{11}$              \\
$\sigma_{\rm RV,SOPHIE}$ (m/s)       & $10.3^{+13.0}_{-7.2}$                 & $13.2^{+22.0}_{-9.5}$             \\
$\sigma_{\rm RV,CAFE}$ (m/s)         & -                                & 43 $\pm$ 21           \\
$\Delta$ RV$_{S,H-N}$ (m/s)          & -121.5 $\pm$ 9.1                    & -81 $\pm$ 26           \\
$\Delta$ RV$_{S,CAFE}$ (m/s)         & -                                & 228 $\pm$ 50            \\
$\sigma_{\rm K2}$(ppm)               & 385 $\pm$ 18                     & 188.3 $\pm$ 7.3          \\
$\sigma_{\rm SED}$ (mag)             & $0.016^{+0.016}_{-0.011}$        & $0.065^{+0.032}_{-0.021}$            \\    
K2 contamination                     & $0.032^{+0.036}_{-0.023}$        &  $0.025^{+0.031}_{-0.018}$     \\  \hline                 

\hline                  
\end{tabular}
\tablefoot{
\tablefoottext{a}{Modified Barycentric Julian Date = BJD-2457000 (days)}
}

\end{table}

\end{document}